\documentclass[prl,twocolumn,preprintnumbers]{revtex4}

\usepackage{times}

\usepackage{bm}
\usepackage{graphicx}
\usepackage{amsbsy}
\usepackage{amsmath}
\usepackage{amsfonts}
\usepackage{amsthm}

\begin{document}

\theoremstyle{plain}
\newtheorem{theorem}{Theorem}
\newtheorem{lemma}[theorem]{Lemma}
\newtheorem{corollary}[theorem]{Corollary}
\newtheorem{proposition}[theorem]{Proposition}
\newtheorem{conjecture}[theorem]{Conjecture}

\theoremstyle{definition}
\newtheorem{definition}[theorem]{Definition}

%\theoremstyle{remark}
%\newtheorem*{remark}{Remark}
%\newtheorem{example}

%\title{Secret sharing with three qubit mixed states generated by noisy quantum channel }
\title{Probabilistic Secret Sharing Through Noisy Quantum Channels}
\author {Satyabrata Adhikari, Indranil Chakrabarty, Pankaj Agrawal}
\email{tapisatya@iopb.res.in, indranil@iopb.res.in, agrawal@iopb.res.in}
\affiliation{Institute of Physics, Sainik School Post,
Bhubaneswar-751005, Orissa, India}

\date{\today}

%\maketitle
\begin{abstract}
In a realistic situation, the secret sharing of classical or quantum information will involve 
the transmission of this information through noisy channels. We consider a three qubit
pure state. This state becomes a mixed-state when the qubits are distributed over
noisy channels.  We focus on a specific noisy channel, the phase-damping channel.
We propose a protocol for secret sharing of classical information with this and
related noisy channels.  This protocol can also be thought of as cooperative superdense 
coding. We also discuss other noisy channels to examine the possibility of 
secret sharing of classical information.

\end{abstract}

\maketitle
\section{I. Introduction}

Quantum entanglement \cite{einstein} plays a pivotal role in understanding the deepest nature of reality. In classical world there is no
counter part of quantum entanglement. Entanglement is a very useful resource in the sense that using entanglement a lot of things 
can be done that 
cannot be done otherwise. Entanglement is also essential for the communication tasks like quantum teleportation \cite{bennett2}, quantum cryptography \cite{gisin} and quantum secret sharing \cite{hillery}.\vskip 0.1cm 

In a secret sharing protocol, one distributes a secret message among a group of people. This is done by allocating a share
of the secret to each of these participants. The beauty of the entire secret sharing process lies in the fact that, if there is a 
dishonest member in the group of participants, he will not be able to find the secret  without the collaboration of other  
members. In other words, the secret can be reconstructed only when a sufficient number of shares are combined together; 
individual shares are of no use.\vskip 0.1cm
%In most of the quantum secret sharing protocols, initially a party encodes the information in a qubit. Then he distributes it among several parties in such a way that neither of them is able to decode the entire information individually. However, one of them is able to unmask the secret perfectly, when all the other parties agree to cooperate.
The secret sharing protocol in a quantum scenario was first introduced in 
Ref \cite{hillery}. After its introduction, Karlsson et.al.\cite{karlsson} studied the similar quantum secret sharing protocol using bipartite pure entangled state. Many authors studied the concept of quantum secret sharing using tripartite pure entangled states and also for multi partite states like graph states \cite{bandyopadhyay,bagherinezhad,lance,gordon,zheng, markham}. Recently Q. Li et.al. proposed semi-quantum secret sharing protocols using maximally entangled GHZ state which
was shown to be secure against eavesdropping  \cite{li}. Recently in  \cite{satya}, it was shown that Quantum secret sharing is possible with bipartite two qubit mixed states (formed due to noisy environment). Quantum secret sharing can also be realized in experiment \cite{tittel, schmid, schmid1, bogdanski}.\vskip 0.1cm

The purpose of this paper is to introduce a protocol which can be used to secretly  share classical information in 
the presence of noisy quantum communication channels. We first show that this secret sharing scheme is 
deterministically possible for a shared pure three qubit GHZ state. Then,
we consider a realistic situation where a source creates a pure GHZ state and then the qubits are distributed 
to different parties through noisy channels. These noisy channels convert the initial pure state into a mixed state. 
We carry out the analysis and find the number of classical bits that can be secretly shared for a specific 
noisy channel, the phase-damping channel. One of the important feature 
of this channel is that it describes the loss of quantum information
without loss of energy. We also talk about several other noisy channels and comment on  the possibility of 
secret sharing using those channels.

%This is because of the effects of the noisy channel on the transferred qubits. %In case of phase-flip channel the secret sharing is not perfectly %possible. Interestingly, we find that in case of phase damping channel our secret sharing protocol is successful deterministically while it is successful %probabilistically if phase-flip channel, amplitude damping channel etc. are employed.\vskip 0.1cm

 The organization of the paper is as follows. In Section II, we describe our protocol for pure three qubit GHZ state. In Section III, we deviate from the 
ideal scenario and consider the realistic situation where qubits are transferred through phase-damping channels and reinvestigate our secret sharing scheme. In the last section, we discuss other noisy channels
and present our conclusions.

\section{II. Secret sharing scheme with shared pure GHZ state}
In this section, we introduce a protocol for quantum secret sharing with shared pure GHZ state. In this protocol, three parties start with a shared pure GHZ state. Then one of the members encodes secret by doing some local unitary operation on her qubit. Thereafter, she sends her qubit to one of the other two members. Interestingly neither of these two  members would be able to know about the local unitaries performed by the encoder individually. However, we show that if they agree to collaborate, then one of the parties can decode
the two bit secrets. Our protocol goes like this.\\

\noindent\textbf{Step I: Pure GHZ State shared by three parties}\vskip 0.1cm
\noindent Let us consider three parties say, Alice (A), Bob (B) and Charlie (C) share a pure GHZ state
\begin{eqnarray}
|\Psi\rangle_{ABC}=\frac{1}{\sqrt{2}}[|000\rangle+|111\rangle]. \label{ghz} 
\end{eqnarray}
\noindent\textbf{Step II: Unitary operations performed by Alice}\vskip 0.1cm
\noindent In this step, Alice encodes two bits of secret information by performing one of the $\{I, \sigma_x, i \sigma_y, \sigma_z\}$   unitary operations on her qubit. After performing one of the unitary operations the state (\ref{ghz}) transforms correspondingly to one of the following states
\begin{eqnarray}
(I \otimes I \otimes I)|\Psi\rangle_{ABC} =\frac{1}{\sqrt{2}}[|000\rangle+|111\rangle],\nonumber\\ \label{unitary}
(\sigma_x \otimes I \otimes I)|\Psi\rangle_{ABC} =\frac{1}{\sqrt{2}}[|100\rangle+|011\rangle],\nonumber\\ 
(i \sigma_y \otimes I \otimes I)|\Psi\rangle_{ABC} =\frac{1}{\sqrt{2}}[|100\rangle-|011\rangle], \nonumber\\
(\sigma_z \otimes I \otimes I)|\Psi\rangle_{ABC} =\frac{1}{\sqrt{2}}[|000\rangle-|111\rangle].
\end{eqnarray}
Alice then sends her qubit to Bob.\vskip 0.1cm

\noindent\textbf{Step III: Charlie performs single-qubit measurement}\vskip 0.1cm

%\noindent After recieving particle from Bob, Charlie now have two particles at his disposal.\vskip 0.1cm

 The above set of equations (\ref{unitary}) can be rewritten as
\begin{eqnarray}
|\Psi\rangle^{I}_{BBC} =\frac{1}{2}[|\Phi^{+}\rangle\otimes(|0\rangle+|1\rangle)\nonumber\\+|\Phi^{-}\rangle\otimes(|0\rangle-|1\rangle)],\nonumber\\
|\Psi\rangle^{X}_{BBC} =\frac{1}{2}[|\Psi^{+}\rangle\otimes(|0\rangle+|1\rangle)\nonumber\\-|\Psi^{-}\rangle\otimes(|0\rangle-|1\rangle)],\nonumber\\
|\Psi\rangle^{Y}_{BBC} =\frac{1}{2}[|\Psi^{+}\rangle\otimes(|0\rangle-|1\rangle)\nonumber\\-|\Psi^{-}\rangle\otimes(|0\rangle +|1\rangle)],\nonumber\\
|\Psi\rangle^{Z}_{BBC} =\frac{1}{2}[|\Phi^{+}\rangle\otimes(|0\rangle-|1\rangle)\nonumber\\+|\Phi^{-}\rangle\otimes(|0\rangle+|1\rangle)],
\end{eqnarray}
where $|\Phi^{\pm}\rangle = \frac{1}{\sqrt{2}}[|00\rangle\pm|11\rangle], |\Psi^{\pm}\rangle = \frac{1}{\sqrt{2}}[|01\rangle\pm|10\rangle]$.\vskip 0.1cm

At this stage, it is not possible either for Bob or for Charlie to decipher the secret encoded by Alice. However, Bob can unmask the secret if Charlie agrees to cooperate with him. Since Charlie now has a single particle at his disposal, he performs a single-qubit measurement in the Hadamard basis $\{\frac{|0\rangle+|1\rangle}{\sqrt{2}}, \frac{|0\rangle-|1\rangle}{\sqrt{2}}\}$. Then he can help Bob to decode the message by conveying to him the outcomes of his measurement. \vskip 0.1cm

\noindent\textbf{Step IV: Bob performs Bell-state measurement}\vskip 0.1cm
According to the measurement outcomes announced by Charlie, Bob performs a Bell-state measurement on his two qubits. According to his Bell-state measurement outcome, he  can find the secret encoded by Alice. The two bits secret decoded by Bob as a result of the declaration of the measurement outcome by Charlie is given in the following table.\\\vskip 0.1cm
 
%~~~~~~~~~~~~~~~~~~~~~~~~~~~~~~~~~~~~~\underline{{\bf TABLE I:}}\\
\begin{tabular}{|c|c|c|}
\hline  Qubits at  & Charlie's Measurement  & Secrets deciphered \\ Bob's side  & Outcome & by Bob\\
\hline $|\Phi^{+}\rangle_{BB}$ & $\frac{1}{\sqrt{2}}[|0\rangle_{C}+|1\rangle_{C}]$ & $I$\\
 & $\frac{1}{\sqrt{2}}[|0\rangle_{C}-|1\rangle_{C}]$ & $\sigma_z$  \\
\hline $|\Phi^{-}\rangle_{BB}$ & $\frac{1}{\sqrt{2}}[|0\rangle_{C}+|1\rangle_{C}]$ &  $\sigma_z$ \\
& $\frac{1}{\sqrt{2}}[|0\rangle_{C}-|1\rangle_{C}]$  & $I$ \\
\hline $|\Psi^{+}\rangle_{BB}$ & $\frac{1}{\sqrt{2}}[|0\rangle_{C}+|1\rangle_{C}]$ &  $\sigma_x$ \\
& $\frac{1}{\sqrt{2}}[|0\rangle_{C}-|1\rangle_{C}]$ & $i \sigma_y$ \\
\hline $|\Psi^{-}\rangle_{BB}$ & $\frac{1}{\sqrt{2}}[|0\rangle_{C}+|1\rangle_{C}]$ &  $i \sigma_y$  \\
& $\frac{1}{\sqrt{2}}[|0\rangle_{C}-|1\rangle_{C}]$  &  $\sigma_x$ \\
\hline
\end{tabular}\\\\
\section{III. Secret sharing with mixed state}
 In this section, we consider a more realistic situation in which a party, say, Charlie generates a three qubit pure GHZ state in his laboratory. Then he 
keeps one qubit with him and sends the other two qubits through two identical noisy channels to two of his friends Alice (A) and Bob (B). As a result of the interaction of the qubit with the environment, the three parties share a three-qubit mixed state after
the distribution of the qubits. We would consider the case of phase-damping channel as the noisy channel.
The action of a phase-damping channel are given by the set of three Kraus operators $E_0=\sqrt{1-p}I, E_1=\sqrt{p}|0\rangle\langle 0|, E_2=\sqrt{p}|1\rangle\langle 1|$, where $p~ (0<p<1)$ is the channel parameter \cite{ting}. In this case, our protocol for secret sharing may be described in the following steps:\vskip 0.1cm

\noindent\textbf{Step I: Transferring qubits through phase-damping channels}\vskip 0.1cm
 Let us assume that Charlie prepares a three-qubit pure GHZ state. Thereafter, he keeps one qubit with him and sends the other two qubits through two 
phase-damping channels to two of his friends Alice and Bob. As a result of the action of the channels, described by the above Kraus operators, on their respective qubits, the resultant state shared by the three parties becomes a  mixed state
\begin{eqnarray}
\rho^{ABC}=\frac{1}{2}[|000\rangle\langle 000|+ |111\rangle\langle 111|]\nonumber\\+\frac{(1-p)^2}{2}[|000\rangle\langle 111|+ |111\rangle\langle 000|].
\end{eqnarray}
\noindent\textbf{Step II: Local Unitary operation by Alice }\vskip 0.1cm
 After receiving the qubits from Charlie, one of the party say Alice encodes two bits of information by carrying out the local unitary 
transforms $\{I(00),\sigma_x (01), i \sigma_y (10), \sigma_z (11)\}$ on her qubit. After performing one of the unitary operations the state  
transforms to one of the following states
\begin{eqnarray}
I:\rho_I^{ABC}=\frac{1}{2}[|000\rangle\langle 000|+ |111\rangle\langle 111|]\nonumber\\+\frac{(1-p)^2}{2}[|000\rangle\langle 111|+ |111\rangle\langle 000|],\nonumber \\
\sigma_x: \rho_X^{ABC}=\frac{1}{2}[|100\rangle\langle 100|+ |011\rangle\langle 011|]\nonumber\\+\frac{(1-p)^2}{2}[|100\rangle\langle 011|+ |011\rangle\langle 100|],\nonumber \\
i \sigma_y: \rho_Y^{ABC}=\frac{1}{2}[|100\rangle\langle 100|+ |011\rangle\langle 011|]\nonumber\\-\frac{(1-p)^2}{2}[|100\rangle\langle 011|+ |011\rangle\langle 100|],\nonumber \\
\sigma_z:\rho_Z^{ABC}=\frac{1}{2}[|000\rangle\langle 000|+ |111\rangle\langle 111|]\nonumber\\-\frac{(1-p)^2}{2}[|000\rangle\langle 111|+ |111\rangle\langle 000|].
\end{eqnarray}

Then Alice sends her qubit to Bob through the same phase-damping channel,  described by the channel parameter $p$. Bob now has two qubits with him while the third qubit is with Charlie.
As a result, the three-qubit density operators representing the above states  reduce to the states

% These four possible three qubit states corrosponding to four unitary operations %done by Alice are given by
\begin{eqnarray}
\rho_1^{BBC}=\frac{1}{2}[|000\rangle\langle 000|+ |111\rangle\langle 111|]\nonumber\\+\frac{(1-p)^3}{2}[|000\rangle\langle 111|+ |111\rangle\langle 000|], \nonumber \\
\rho_2^{BBC}=\frac{1}{2}[|100\rangle\langle 100|+ |011\rangle\langle 011|]\nonumber\\+\frac{(1-p)^3}{2}[|100\rangle\langle 011|+ |011\rangle\langle 100|], \nonumber \\
\rho_3^{BBC}=\frac{1}{2}[|100\rangle\langle 100|+ |011\rangle\langle 011|]\nonumber\\-\frac{(1-p)^3}{2}[|100\rangle\langle 011|+ |011\rangle\langle 100|],\nonumber \\
\rho_4^{BBC}=\frac{1}{2}[|000\rangle\langle 000|+ |111\rangle\langle 111|]\nonumber\\-\frac{(1-p)^3}{2}[|000\rangle\langle 111|+ |111\rangle\langle 000|].
\end{eqnarray}

 \vskip 0.2cm

\noindent\textbf{Step III: Charlie performs single qubit measurement}\vskip 0.1cm

 After receiving the particle from Alice, Bob has two particles at his disposal. But, it is not possible either for Bob or for Charlie 
independently to decode the information encoded by Alice. However, Bob can decode the secret if Charlie is willing to help him. Charlie cooperates by conveying his measurement outcomes. Charlie performs a measurement on his qubits in the basis $\{|+\rangle=\alpha|0\rangle+\beta|1\rangle, |-\rangle=\beta|0\rangle-\alpha|1\rangle \}$, (where $\alpha^2+\beta^2=1$). As a result of this measurement, the state that collapses on Bob's side are given by the following table \\

%~~~~~~~~~~~~~~~~~~~~~~~~~~~~~~~~~~~~~\underline{{\bf TABLE II:}}\\
\begin{tabular}{|c|c|c|c|}
\hline Secret & State shared by  & Charlie's Measurement & Qubits at\\  Encoded & Bob and Charlie  & Outcomes & Bob's side \\
\hline $I(00)$  &$\rho_1^{BBC}$ & $|+\rangle$ & $\rho_1^{BB+}$ \\
&& $|-\rangle$ & $\rho_1^{BB-}$ \\
\hline $\sigma_x (01)$ & $\rho_2^{BBC}$ & $|+\rangle$ & $\rho_2^{BB+}$ \\
 && $|-\rangle$ & $\rho_2^{BB-}$ \\
\hline $i \sigma_y (10)$ & $\rho_3^{BBC}$ & $|+\rangle$ & $\rho_3^{BB+}$ \\
 && $|-\rangle$ & $\rho_3^{BB-}$ \\
\hline $\sigma_z (11)$ & $\rho_4^{BBC}$ & $|+\rangle$ & $\rho_4^{BB+}$ \\
 && $|-\rangle$ & $\rho_4^{BB-}$ \\
\hline
\end{tabular} \\
\\
where,
\begin{eqnarray}
 \rho_1^{BB+}=&&\alpha^2|00\rangle\langle 00|+\beta^2|11\rangle\langle 11|{}\nonumber\\&&+\alpha\beta(1-p)^3(|11\rangle\langle 
00|+|00\rangle\langle 
11|),\nonumber\\
\rho_1^{BB-}=&&\beta^2|00\rangle\langle 00|+\alpha^2|11\rangle\langle 11|{}\nonumber\\&&-\alpha\beta(1-p)^3(|11\rangle\langle 00|+|00\rangle\langle 11|),\nonumber\\
\rho_2^{BB+}=&&\alpha^2|10\rangle\langle 10|+\beta^2|01\rangle\langle 01|{}\nonumber\\&&+\alpha\beta(1-p)^3(|01\rangle\langle 10|+|10\rangle\langle 01|),\nonumber\\
\rho_2^{BB-}=&&\beta^2|10\rangle\langle 10|+\alpha^2|01\rangle\langle 01|{}\nonumber\\&&-\alpha\beta(1-p)^3(|01\rangle\langle 10|+|10\rangle\langle 01|),\nonumber\\
\rho_3^{BB+}=&&\alpha^2|10\rangle\langle 10|+\beta^2|01\rangle\langle 01|{}\nonumber\\&&-\alpha\beta(1-p)^3(|01\rangle\langle 10|+|10\rangle\langle 01|),\nonumber\\
\rho_3^{BB-}=&&\beta^2|10\rangle\langle 10|+\alpha^2|01\rangle\langle 01|{}\nonumber\\&&+\alpha\beta(1-p)^3(|01\rangle\langle 10|+|10\rangle\langle 01|),\nonumber\\
 \rho_4^{BB+}=&&\alpha^2|00\rangle\langle 00|+\beta^2|11\rangle\langle 11|{}\nonumber\\&&-\alpha\beta(1-p)^3(|11\rangle\langle 
00|+|00\rangle\langle 11|),\nonumber\\ 
\rho_4^{BB-}=&&\beta^2|00\rangle\langle 00|+\alpha^2|11\rangle\langle 11|{}\nonumber\\&&+\alpha\beta(1-p)^3(|11\rangle\langle 00|+|00\rangle\langle 11|).
\end{eqnarray}
Charlie sends his results to Bob through a classical channel by spending one classical bit. This is done by encoding $0$ for $|+\rangle$ and $1$ for $|-\rangle$ respectively.

\vskip 0.2cm 

\noindent\textbf{Step IV: Bob performs two qubit projective measurement and POVM}\vskip 0.1cm

As we see in the above table when Charlie's measurement result is $|+\rangle$ then Bob can have one of the four possible states $\rho_1^{BB+},\rho_2^{BB+},\rho_3^{BB+},\rho_4^{BB+}$. Similarly when Charlie's qubit collapses into the state  $|-\rangle$, Bob can have any one of the state four possible states $\rho_1^{BB-},\rho_2^{BB-},\rho_3^{BB-},\rho_4^{BB-}$ at his disposal.

If Charlie sends $0$, then Bob guesses that the two qubit states in his possession would be either $\rho_1^{BB+}$ or $\rho_2^{BB+}$ or $\rho_3^{BB+}$ or $\rho_4^{BB+}$. He then performs projective measurements $P_1=|00\rangle\langle 00|+|11\rangle\langle 11|$ and $P_2=|01\rangle\langle 01|+|10\rangle\langle 10|$ to get close to identify the secret. The projectors $P_1$ and $P_2$ classify the above four states into two classes as  $C_1=\{\rho_1^{BB+}, \rho_4^{BB+}\}$ and $C_2=\{\rho_2^{BB+}, \rho_3^{BB+}\}$ respectively. The states within the two classes are now lying in a two dimensional subspace spanned by $\{|00\rangle, |11\rangle \}$ and $\{|01\rangle, |10\rangle \}$ respectively.\\

After classifying the states,  Bob performs optimal POVM  to identify the state in which secret is encoded. First of all he considers the class $C_1=\{\rho_1^{BB+},\rho_4^{BB+}\}$ and constructs the optimal POVM operators $\Pi_1$, $\Pi_2$ for discriminating the density matrices present in the class.
The optimal POVM  measurement  is the one that minimizes the error rate
\begin{eqnarray}
E_R=\frac{1}{2}[Tr[\Pi_1\rho_4^{BB+}]+Tr[\Pi_2\rho_1^{BB+}].
\end{eqnarray}
subject to the constraints that they forms a complete set of projectors (i.e $\Pi_1+\Pi_2=I$) \cite{mj}. 

The optimal POVM elements are

\begin{eqnarray}
\Pi_{1}=\frac{1}{2}\left(%
\begin{array}{cc}
  1 & 1  \\
  1 & 1 \\
\end{array}%
\right),
\Pi_{2}=\frac{1}{2}\left(%
\begin{array}{cc}
  1 & -1  \\
  -1 & 1 \\
\end{array}%
\right).
\end{eqnarray}
and the error rate in discriminating the states $\rho_1^{BB+}$ and $\rho_4^{BB+}$ is \cite{mj}
\begin{eqnarray}
E_R=\frac{1}{2}(1-2\alpha\beta(1-p)^3).
\end{eqnarray}

Similarly, Bob can distinguish the mixed states belonging to the other class $C_2=\{\rho_2^{BB+}, \rho_3^{BB+}\}$ 
by using the same set of POVM operators $\Pi_1$ and $\Pi_2$. The error rate $E_R$ in this case will also be the same.
Therefore the total probability of success in distinguishing these states is

\begin{eqnarray}
P_S= 2\alpha\beta(1-p)^3. 
\end{eqnarray}

%                          FIGURE 1
\begin{figure}[t]
\begin{center}

\[
\begin{array}{cc}
\includegraphics[height=6.0cm,width=6cm]{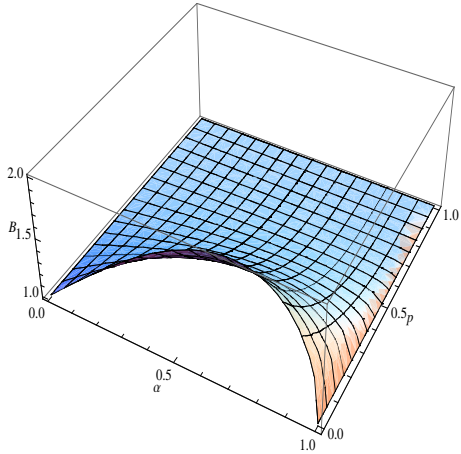}
\end{array}
\]
\caption{ $B$ (the classical bits decoded by Bob) is plotted against the parameters $\alpha$ and $p$}

\end{center}
\end{figure}

\begin{figure}[t]
\begin{center}

\[
\begin{array}{cc}
\includegraphics[height=6.0cm,width=6cm]{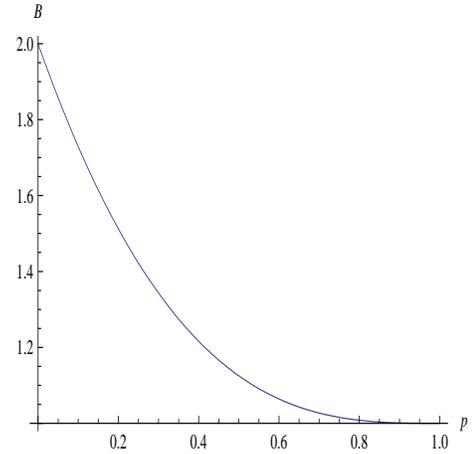}
\end{array}
\]
\caption{ $B$ (the classical bits decoded by Bob) is plotted against the channel parameter $p$ when the measurement is done in the Hadamard basis}

\end{center}
\end{figure}

In such a situation, the  total number of classical bits that Bob can extract are

\begin{eqnarray}
B = 1+2\alpha\beta(1-p)^3. 
\end{eqnarray}

Clearly, the amount of classical information that can be extracted by Bob will depend upon the channel noise ($p$) 
and also on the basis that Charlie uses for the measurement. We note that when $p=1$, the channel is totally noisy, and
Bob can extract at most one classical bit. This can also be seen from Figure 1.  $B$ is independent of $\alpha$ and is
always equal to $1$ when $p=1$. The limit $p=0$ corresponds to the case when there is no noise. In this case,  $B$ is maximum 
when measurement has been done in the Hadamard basis (i.e $\alpha=\beta=\frac{1}{\sqrt{2}}$). 
This is also clear from the Figure 1. In general, the $B$ obtains the largest value when Charlie makes his measurement
in Hadamard basis. In such a scenario

\begin{eqnarray}
B = 1+ (1-p)^3.
\end{eqnarray}

 In Figure 2, we have plotted $B$ as a function of the channel parameter ($p$).  It takes maximum value 2 when $p=0$ 
 and the minimum value $1$ when $p=1$.\\

Thus we see that deterministic  secret sharing is not  possible with a phase-damping channel. We also find that the 
amount of classical information decoded  by Bob is dependent on the noise parameter ($p$) and also on the choice of basis.
 In a practical situation when we carry out quantum information processing task we face the decoherence problem and we always have mixed state at our disposal. As a consequence of which the tasks which can be done deterministically in case of pure states, can not be done so for the mixed states. 

\section{IV. Discussion and Conclusions}
In this paper, we have introduced a protocol for secret sharing which is different from the existing secret sharing
 schemes. We have considered a realistic scenario where there are noisy quantum channels. In such a scenario, 
 the deterministic secret sharing is not possible. We consider POVM measurements to implement our protocol and
 find out the number of classical bits that Alice can share with Bob with the help of Charlie. The answer is $ 1 + (1-p)^3$
 classical bits with $p$ characterizing the noisy channel. The earlier scheme of secret sharing with pure GHZ state 
 was more like cooperative teleportation while our scheme of secret sharing is like cooperative dense coding. 
 The three-qubit mixed state considered here is generated by passing the qubits through noisy channels. 
 In particular, we have shown how the phase-damping noisy channel  generated three-qubit mixed state can be used in our 
  secret sharing protocol.\vskip0.1cm

 Now it would be important to ask  that whether our secret sharing scheme succeeds only when the noisy channel
is a phase-damping channel. Indeed the answer is 'no'. We find that if phase-flip channel is the noisy channel, 
then our secret sharing scheme would succeed. However, one needs to explore further if the secret sharing 
scheme can succeed with noisy channels like amplitude-damping channel, depolarizing channel, bit-flip channel, 
bit-phase flip channel or two Pauli channels. In the case of phase-damping and phase-flip channels, the Kraus 
operators are diagonal and it is not difficult to construct  appropriate POVM operators. We also note that these 
channels are related by unitary transformations. Therefore, it appears that  our proposed protocol would 
succeed if the noisy channel is related to phase-damping channel by a unitary transformation.
The reason behind the success of our protocol may be the diagonal form of the Kraus operators 
that represent the noisy channels.  The cases of other noisy channels that are described by the Kraus 
operators with off-diagonal elements may involve
loss of energy and need further exploration. In these cases, probabilistic secret sharing may be possible
with more complicated POVM measurements.

\textit{Acknowledgement}: The authors acknowledge Dr. R. Srikanth and Dr. S. Bandopadhyay for having useful discussions.

\end{document}